# Clear variation of spin splitting by changing electron distribution at non-magnetic metal/Bi$_2$O$_3$ interfaces

H. Tsai[1], S. Karube[1], K. Kondou[2*], N. Yamaguchi[3], F. Ishii[4] and Y. Otani[1, 2,*)]

1 Institute for Solid State Physics, University of Tokyo, Kashiwa 277-8581, Japan
2 Center for Emergent Matter Science, RIKEN, 2-1 Hirosawa, Wako 351-0198, Japan
3 Division of Mathematical and Physical Sciences, Graduate School of Natural Science and Technology, Kanazawa University, Kanazawa 920-1192, Japan
4 Faculty of Mathematics and Physics, Institute of Science and Engineering, Kanazawa University, Kanazawa 920-1192, Japan.

Large spin splitting at Rashba interface, giving rise to strong spin-momentum locking, is essential for efficient spin-to-charge conversion. Recently, a Cu/Bismuth oxide (Bi$_2$O$_3$) interface has been found to exhibit an efficient spin-to-charge conversion similar to a Ag/Bi interface with large Rashba spin splitting. However, the guiding principle of designing the metal/oxide interface for the efficient conversion has not been clarified yet. Here we report strong non-magnetic (NM) material dependence of spin splitting at NM/Bi$_2$O$_3$ interfaces. We employed spin pumping technique to inject spin current into the interface and evaluated the magnitude of interfacial spin-to-charge conversion. We observed large modulation and sign change in conversion coefficient which corresponds to the variation of spin splitting. Our experimental results together with first-principles calculations indicate that such large variation is caused by material dependent electron distribution near the interface. The results suggest that control of interfacial electron distribution by tuning the difference in work function across the interface may be an effective way to tune the magnitude and sign of spin-to-charge conversion and Rashba parameter at interface.

(173 words)

___________________________

*Corresponding authors: kkondou@riken.jp, yotani@issp.u-tokyo.ac.jp

Rashba interface, that has a momentum-dependent spin splitting due to atomic spin-orbit coupling (SOC) and broken inversion symmetry at the interface, plays a key role in spintronics [1-2]. Recently, the Rashba interface has been employed for efficient spin-charge (S-C) current interconversion [3, 4]. The conversion efficiency between spin and charge currents can be comparable or even larger than typical spin Hall materials such as Pt and W [5]. Thus, Rashba effect has been studied intensively as an alternative phenomenon of spin Hall effect (SHE) to control the magnetization by spin current in spintronics devices [6, 7].

Figure 1a shows the Rashba spin splitting in x-y plane, of which Rashba Hamiltonian can be described as $\boldsymbol{H}_R = \alpha_R (\boldsymbol{p} \times \hat{\boldsymbol{z}}) \cdot \boldsymbol{\sigma}$; where $\boldsymbol{\sigma}$ is the vector of Pauli spin matrices, $\boldsymbol{p}$ is the momentum, and $\alpha_R$ is so-called Rashba parameter which determines the splitting in momentum between spin-up and spin-down electrons. The conduction electron spins are aligned to the fictitious field along $\boldsymbol{p} \times \hat{\boldsymbol{z}}$ direction, forming a clockwise or counterclockwise spin texture. Flow of the charge current in the Rashba interface thus generates non-equilibrium spin accumulation, whose gradient drives a diffusive spin current into an adjacent conductive layer. This charge-to-spin (C-S) conversion is called the direct Edelstein effect (DEE). In reverse, injecting the spin current into the interface generates charge current via the interfacial Rashba effect. This phenomenon is called the inverse Edelstein effect (IEE), which has recently been demonstrated using Ag(111)/Bi interface with large Rashba splitting [3].

More recently, we found the similar S-C conversion at the Cu/Bismuth oxide ($Bi_2O_3$) interface by means of several techniques [8-10]. The experimental results revealed the presence of large spin splitting at the Cu/$Bi_2O_3$ interface. In order to obtain more efficient S-C conversion, it is worth understanding how to tune the spin splitting at this metal/oxide type interface.

The Rashba parameter $\alpha_R$ can be described as [11]

$$\alpha_R = \left(\frac{2}{c^2}\right) \int (\partial V / \partial z) |\psi|^2 \, dz, \tag{1}$$

where $c$, $\partial V / \partial z$ and $|\psi|^2$ are respectively the speed of light, potential gradient and electron density distribution. $z = 0$ at the center of atoms at interface. Figure 1b shows a schematic illustration of $V$ and $|\psi|^2$ at NM/$Bi_2O_3$ interfaces based on our ab-initio calculation. Most of the electrons are localized near the NM nuclei because of less charge density in the insulating $Bi_2O_3$ layer than the conductive NM layer. The potential gradient $\partial V / \partial z$ in the vicinity of nuclei is dominant by the antisymmetric Coulomb force of the nucleus as shown in Figure 1b [12, 13]; electron density distribution $|\psi|^2$ is determined by the hybridization state at the interface. Because the integral in equation (1) is strongly affected by asymmetric feature of $|\psi|^2$ [12, 13], even a small modulation of $|\psi|^2$ can have notable effect on $\alpha_R$, i.e. tuning Rashba spin splitting by changing surface potential [14]. This suggests that the Rashba spin splitting can be controlled effectively by tuning the interfacial condition. In this study, we investigated the S-C conversion and Rashba parameter in various NM/$Bi_2O_3$ interfaces and demonstrate the clear variation of Rashba spin splitting by changing electron distribution.

**Experimental results**

**Detection of spin-to-charge conversion in NM (Ag, Cu, Au, Al) /Bi$_2$O$_3$ interfaces.** Fig 1(c) is a schematic illustration of the measurement setup. We prepared four different NM material samples. Each Ni$_{80}$Fe$_{20}$ (Py: 5 nm)/NM (Ag, Cu, Au, or Al 20 nm)/ Bi$_2$O$_3$ (30 nm) tri-layer wire is placed beside a signal line of coplanar waveguide (CPW). The measured samples are fabricated by using photo-lithography and e-beam evaporation (see Method). The length and width of the wire are 200 μm and 14 μm, respectively.

Fig 1(d) is the schematic of spin-to-charge conversion at the NM/Bi$_2$O$_3$ interface. Ferromagnetic resonance (FMR) in Py layer is excited by rf current generated magnetic field $h_{rf}$ in the CPW. Spin current caused by FMR is injected into NM/Bi$_2$O$_3$ layer. This spin current gives rise to an electric dc voltage $V$ through the inverse spin Hall effect (ISHE) and/or inverse Edelstein effect (IEE). All measurements were performed at room temperature. The measurement results are shown in Fig. 2. Clear signals due to spin-to-charge (S-C) conversion are detected for all samples. At the vertical axis, we show the output current values estimated from $V$ because the sample resistance $R$ is different in each sample. The angle $\theta$ is the angle between sample wire and external magnetic field $H$ as shown in Fig 1(c). From this measurement, a strong NM materials dependence in amplitude and sign of detected signals is observed. The signal amplitude is almost the same between Py/Cu/Bi$_2$O$_3$ and Py/Ag/Bi$_2$O$_3$, but surprisingly their signs are opposite each other. While the amplitude of Py/Au(Al)/Bi$_2$O$_3$ is one order or two orders of magnitude smaller than Cu/Bi$_2$O$_3$.

The contribution of ISHE in both Cu and Ag layers can be neglected since the values of spin Hall angle (SHA) for Cu and Ag are too small to explain the detected voltages [3, 8] (see section 1 in supplementary information). The possibility of Bi impurity induced extrinsic spin Hall effect in NM can be excluded because the SH angles induced by Bi in Cu and Ag are both negative [15]. Therefore, the influence of Bi impurities cannot explain the sign change of S-C conversion between Ag/Bi$_2$O$_3$ and Cu/Bi$_2$O$_3$. In addition, there is no notable difference between resistivities of Cu/Al$_2$O$_3$ and Cu/Bi$_2$O$_3$ bilayers, indicating that the contribution of Bi impurities should be small, and the S-C conversions in Py/(Cu, Ag)/Bi$_2$O$_3$ are dominated by IEE at their (Cu, Ag)/Bi$_2$O$_3$ interfaces.

While the contribution of ISHE in Au may be notable since SHA of Au is one order of magnitude lager than Cu and Ag [16, 17]. To estimate the contribution of ISHE in Au, we prepared the reference sample of Py/Au/Al$_2$O$_3$ trilayer. Figure 2(c) shows the output spectrum of Py/Au/Al$_2$O$_3$ and Py/Au/Bi$_2$O$_3$. From the signal amplitude in Py/Au/Al$_2$O$_3$, we estimated spin Hall angle $\theta_{SH}$ in Au layer is +0.40±0.07% (see section 1 in supplementary information), which is in good agreement with reported values [17, 18]. By comparing the signal amplitudes of Py/Au/Al$_2$O$_3$ and Py/Au/Bi$_2$O$_3$, we found that the sign of S-C conversion at Au/Bi$_2$O$_3$ interface should be opposite to SHA in Au.

The rf power-dependence of 5 samples is shown in the upper insets to Fig. 2(a)-(d). The detected signals increase linearly with the rf power, being consistent with the prediction of spin pumping model [19]; It also indicates that the spin pumping experiment are in the linear regime of FMR. Furthermore, the angular dependence of the normalized signal is shown in the lower insets to Fig. 2(a)-(d). All of them show the sinusoidal shape which is consistent with typical IEE model for 2D electron gas. This confirms that the observed S-C conversion signals arise from FMR spin pumping.

**Spin-to-charge conversion coefficient and effective Rashba parameter in NM/Bi$_2$O$_3$ interfaces.**
Table 1 shows the conversion coefficient $\lambda_{\text{IEE}}$, effective Rashba parameter $\alpha_\text{R}^{\text{eff}}$, $|\alpha_\text{R}|$ estimated from first-principle calculation, damping constant $\delta_{\text{eff}}$, and spin mixing conductance $g_{\text{eff}}^{\uparrow\downarrow}$ of different NM/Bi$_2$O$_3$ interfaces. Spin current density injected into NM/Bi$_2$O$_3$ interface is given by [20]

$$J_{\text{s(NM/Bi}_2\text{O}_3)} = \frac{2e}{\hbar} \times \frac{\hbar g_{\text{eff}}^{\uparrow\downarrow} \gamma_e^2 (\mu_0 h_{\text{rf}})^2 \left[\mu_0 M_s \gamma_e + \sqrt{(\mu_0 M_s)^2 \gamma_e^2 + 4\omega^2}\right]}{8\pi \delta_{\text{F/N/O}}^2 [(\mu_0 M_s)^2 \gamma_e^2 + 4\omega^2]} \times e^{(-\frac{t_N}{\lambda_N})} \quad (2)$$

where $\gamma_e$, $M_s$, $\omega$, $h_{\text{rf}}$, $t_\text{N}$, and $\lambda_\text{N}$ are the gyromagnetic ratio, saturation magnetization, angular frequency, applied rf field, thickness of NM layer, and spin diffusion length of NM, respectively. More detailed experiment and calculation methods for estimation of spin current density is explained in Methods. This spin current is converted to charge current at the interface by IEE. The resulting charge current density $j_c$ flowing in the two-dimensional interface is expressed as $j_c = V/wR$, where $V$, $w$, and $R$ are detected voltage, the width of the sample wire, and total resistance of the wire, respectively. For NM=Ag, Cu, Al, the conversion coefficient $\lambda_{\text{IEE}}$ is calculated by $\lambda_{\text{IEE}} = j_c / J_{\text{s(NM/Bi}_2\text{O}_3)}$. Here, the units of $j_c$ and $J_{\text{s(NM/Bi}_2\text{O}_3)}$ are A/m and A/m$^2$, respectively. Therefore, $\lambda_{\text{IEE}}$ has a unit of length. The estimated $\lambda_{\text{IEE}}$ at NM/Bi$_2$O$_3$ (NM = Cu, Ag) interfaces is comparable with the reported value $\lambda_{\text{IEE}} = $ 0.3 nm for Ag/Bi interface measured by spin pumping method [4], and is one-order larger than $\lambda_{\text{IEE}} = $ 0.009 nm for Cu/Bi measured by lateral spin valves method [21]. For NM=Au case, we separated the contribution of SHE and IEE for estimating $\lambda_{\text{IEE}}$. (see section 1 in supplementary information).

The $\lambda_{\text{IEE}}$ can be expressed by using the Rashba parameter $\alpha_R$ and momentum relaxation time $\tau_e^{\text{int}}$ at the interface [22],

$$\lambda_{\text{IEE}} = \alpha_\text{R} \tau_e^{\text{int}} / \hbar \quad (3)$$

In previous study, we showed that $\tau_e^{\text{int}}$ is governed by the momentum relaxation time $\tau_e$ in the NM layer in contact with Rashba interface. By using $\tau_e$ instead of $\tau_e^{\text{int}}$ from the resistivity of NM layer, $\lambda_{\text{IEE}} = \alpha_\text{R}^{\text{eff}} \tau_e / \hbar$, effective Rashba parameter $\alpha_\text{R}^{\text{eff}}$ was calculated. Table 1 shows the strong NM dependence of $\lambda_{\text{IEE}}$ and $\alpha_\text{R}^{\text{eff}}$ at NM/Bi$_2$O$_3$ interfaces. We found that Cu/Bi$_2$O$_3$ and Ag/Bi$_2$O$_3$ have larger $|\alpha_\text{R}^{\text{eff}}|$ and sign of $\alpha_\text{R}^{\text{eff}}$ at Ag/Bi$_2$O$_3$ is positive while others are negative.

**Table 1 | Conversion coefficient $\lambda_{IEE}$, Rashba parameter $\alpha_R^{eff}$, Damping constant $\delta_{eff}$, and spin mixing conductance $g_{eff}^{\uparrow\downarrow}$.**

| Interface | $\lambda_{IEE}$ (nm) | $\alpha_R^{eff}$ (eV·Å) | $\|\alpha_R\|$ (eV·Å) (calculation) | $\delta_{eff}$ | $g_{eff}^{\uparrow\downarrow}$ ($10^{18}$ m$^{-2}$) |
|---|---|---|---|---|---|
| Ag/Bi$_2$O$_3$ | +0.15 ±0.03 | +0.16 ±0.03 | 0.50 | 0.0168 | 10.78 |
| Cu/Bi$_2$O$_3$ | −0.17 ±0.03 | −0.25 ±0.04 | 0.91 | 0.0154 | 8.27 |
| Au/Bi$_2$O$_3$ | −0.09 ±0.03 | −0.10 ±0.04 | 0.29 | 0.0142 | 3.77 |
| Al/Bi$_2$O$_3$ | −0.01 ±0.002 | −0.055 ±0.011 | ------- | 0.0133 | 4.49 |

**Table 1|** Conversion coefficient $\lambda_{IEE}$, Rashba parameter $\alpha_R^{eff}$, damping constant $\delta_{eff}$, and spin mixing conductance $g_{eff}^{\uparrow\downarrow}$ in various NM/Bi$_2$O$_3$ interfaces.

**First-principles calculations.**

The details of electronic state such as charge density and electrostatic potential at NM/Bi$_2$O$_3$ interface were investigated by first-principles calculations. Figures 3(a)-(b) show the electronic states of the NM(111)/α-Bi$_2$O$_3$ interfaces of which local crystallographic configuration is similar to that of our sample (see Figure S1 in supplementary information). The in-plane length of unit cell is based on the experimental lattice constant of each NM. We also assumed other local crystallographic configuration for the NM/Bi$_2$O$_3$ interfaces in terms of the out of plane arrangement of NM and the crystal phases of Bi$_2$O$_3$ (e.g. NM(110)/β-Bi$_2$O$_3$). The calculated $\alpha_R$ is in the same order of magnitude for both interfaces. From our thickness dependence calculation, we found that the electronic structures were insensitive to the number of NM layers once the number of layers exceeds 16. The value of $\alpha_R$ can be determined from the calculated band structure of each NM(111)/α-Bi$_2$O$_3$ interface (see Figure S3 in supplementary information). The calculated $|\alpha_R|$ in NM(111)/α-Bi$_2$O$_3$ interface are shown in Table 1. The experimental values of $|\alpha_R|$ are about 3 times smaller than the calculated values; this difference may come from the different structure between real samples and the calculations. In the experiment the deposited Bi$_2$O$_3$ layer is amorphous and the NM(111) layer has about 1 nm roughness, so it is reasonable that the smaller $\alpha_R$ is obtained by experiments. The strength dependence of SOC in Bi on the $\alpha_R$ is shown in Fig. 3(c). The $\alpha_R$ without SOC of Bi is in the order of each NM (111) material. For NM = Cu and Ag, the $\alpha_R$ drastically increases as the strength of SOC of Bi increases, while the $\alpha_R$ slightly decreases for NM = Au. The charge density distribution for the corresponding Rashba state $|\psi|^2$ and potential $V$ are shown in Fig. 3(d)-(f). The gradient of potential $\partial V / \partial z$ in NM = Cu is smaller than Ag and Au case, however, $\alpha_R$ of Cu/Bi$_2$O$_3$ is larger than others. This indicates that, in the case of Cu/Bi$_2$O$_3$, $|\psi|^2$ is the dominant essence instead of $\partial V / \partial z$. For NM = Cu and Au, the peak of $|\psi|^2$ shifts to NM side, while for NM = Ag, it shifts to Bi$_2$O$_3$ side. This difference of the asymmetry feature of $|\psi|^2$ may have an influence on the magnitude and, especially, sign of Rashba parameter. In addition, for NM = Cu, the peak of $|\psi|^2$ is strongly localized at the peak of potential, while for NM =

Au, the peak of $|\psi|^2$ becomes broaden; this difference between the localized features may also have an influence on the magnitude of Rashba parameter.

**Discussion**

From the experiments and the first principle calculations, we can confirm that the strong NM dependence of $\alpha_R$ comes from the asymmetric charge density distribution $|\psi|^2$ at interfaces, which is originated from the broken inversion symmetry at interfaces. Besides that, the SOC of the materials is another important essence of Rashba effect. Firstly, we compare the influence of SOC of different NM materials. Even though Au has one order larger SOC than Ag and Cu, its $Bi_2O_3$ interface has smaller $|\alpha_R^{eff}|$. This result suggests that the SOC of NM layer is not essential to Rashba effect at NM/$Bi_2O_3$ interfaces. This trend is the same with the first-principles calculations and experimental results in ARPES measurement in Ag(111)/Bi and Cu(111)/Bi Rashba interfaces [23]. Furthermore, Fig. 3(c) shows that the SOC of Bi dominant the large Rashba spin splitting at NM/$Bi_2O_3$ interface in NM = Ag and Cu cases. Therefore, the strong NM dependency is not due to different SOC strength of NM materials. Secondly, since $|\psi|^2$ should be modulated by the electric field, we discuss here the contribution of interface structure and Fermi energy difference between NM and $Bi_2O_3$ layer which determine the electric field at the interfaces. In the metallic Rashba interface such as Ag/Bi, the interface alloying structure is essential for originating the giant Rashba splitting because it induces strong in-plane potential gradient [24]. For NM/$Bi_2O_3$ interfaces, the value of Rashba parameter at Ag/$Bi_2O_3$ interface is one order smaller than Ag(111)/Bi , and Cu/$Bi_2O_3$ is about half of Cu(111)/Bi [23]. This reduction might be caused by the lack of interface alloying and in-plane potential gradient, because Bi atoms are much more strongly bonded to oxygen atoms than to the NM. In this situation, $\alpha_R$ at NM/$Bi_2O_3$ interface is not only determined by interface alloying structure and the out-of-plane electric field at the interface should become an important essence to induce broken inversion symmetry and the interfacial spin splitting. Since the out-of-plane electric field at the interface originates from work function difference $\Delta\Phi_{NM-Bi2O3}$ (Fermi energy difference) between NM and $Bi_2O_3$, $\alpha_R$ may be related with $\Delta\Phi_{NM-Bi2O3}$. Fig. 4(a) shows absolute value estimated by experiment and calculation in different NM/$Bi_2O_3$ interfaces as a function of $|\Delta\Phi_{NM-Bi2O3}|$. Here, the $\Delta\Phi_{NM-Bi2O3}$ is defined as $\Phi_{NM}-\Phi_{Bi2O3}$. We use reported value of work function $\Phi$ of Cu (111) [25], Ag(111), Au(111), Al(111) [26], and $\alpha$- $Bi_2O_3$ [27] as 4.96, 4.74, 5.31, 4.26, and 4.92 in units of eV, respectively. It seems that $|\alpha_R^{eff}|$ decreases as $|\Delta\Phi_{NM-Bi2O3}|$ increases and the trend of calculated $|\alpha_R|$ is in good agreement with the experimental results.

This trend could be explained by Fig. 1(b), which is supported by the calculation results in Fig. 3(c) and (e). When the interfacial electric field $E_{inter}$, is quite small, the asymmetric $|\psi|^2$ is strongly localized near NM nuclei as shown by purple line. If $E_{inter}$ becomes large enough, the peak of $|\psi|^2$ could be shifted from nuclei and delocalized by charge transfer due to interfacial electric field as shown by blue line. As the result of larger $E_{inter}$, the integral of eq. (1) becomes smaller because $|\psi|^2$ is not localized in the largest potential region, and therefore when $|\Delta\Phi_{\text{NM-Bi2O3}}|$ increases, $|\alpha_R|$ decreases. That is to say, $|\psi|^2$ modulated by interfacial electrical field can drastically change $\alpha_R$. This charge-transfer-induced delocalization of $|\psi|^2$ is often discussed in ferroelectric oxides by Wannier functions [28].

Additionally, we found that there is a sign change of $\alpha_R^{\text{eff}}$ at Ag/Bi$_2$O$_3$ interface as shown in Fig. 4(b). In eq. (1), because the $\partial V / \partial z$ is almost an antisymmetric function with respect to the nucleus, sign of $\alpha_R$ is determined by whether the excess electron density is localized on NM side or Bi$_2$O$_3$ side. The opposite sign between Ag/Bi$_2$O$_3$ and Cu/Bi$_2$O$_3$ should come from the different asymmetry of $|\psi|^2$. When there is a sign change of $\Delta\Phi$, the $E_{\text{inter}}$ in Fig. 1(a) has opposite direction. Assuming that Ag/Bi$_2$O$_3$ and Cu/Bi$_2$O$_3$ interfaces have similar hybridization state, the opposite direction of $E_{\text{inter}}$ may shift the $|\psi|^2$ to different side of NM or Bi$_2$O$_3$ and then cause the sign change of $\alpha_R$. This opposite direction shift is demonstrated by calculation results in Fig. 3(e). Also in case of Gd(0001) and O/Gd(0001) surface, it has been reported that the sign change behavior is caused by asymmetry of $|\psi|^2$ due to top oxide layer [29]. While in case of Al/Bi$_2$O$_3$ interface, the sign is not as expected by the same scenario as NM = Ag, Cu, and Au. Since Al itself has quite different electronic state with Ag, Cu, and Au (group 11 elements), the hybridization state at Al/Bi$_2$O$_3$ interface may have different asymmetric feature with others and that's why Al/Bi$_2$O$_3$ interface does not have the same sign as Ag/Bi$_2$O$_3$ though their $\Delta\Phi_{\text{NM-Bi2O3}}$ are both negative.

In summary, we have demonstrated the large magnitude variation and sign change of S-C conversion originated from Rashba spin-splitting at various NM/Bi$_2$O$_3$ interfaces. This strong variation comes from the material dependent electron distribution near the interface. The experimental results, supported by calculation, suggest that $|\psi|^2$ could be controlled by tuning interfacial electric field between NM and Bi$_2$O$_3$. This study provides a further understanding of the origin of the large spin-splitting at NM/Bi$_2$O$_3$ interfaces, and also shown an effective way to tune the magnitude and sign of S-C conversion by changing the electron distribution. Furthermore, our results and measurement technique may provide a guiding principle for finding novel NM/oxide interfaces with large spin-splitting in the future.

**Methods**

**Sample preparation.** The measured tri-layer samples, Py(5 nm)/NM (Ag, Cu, Au, Al 20 nm)/ Bi$_2$O$_3$ (30 nm), were deposited on SiO$_2$ (200 nm)/Si substrate by e-beam evaporation method. The base pressure in the chamber was $3\times10^{-5}$ Pa. The evaporation rate of Py, NM and Bi$_2$O$_3$ layer were 0.2 Å/s, 2.0 Å/s, and 0.2 Å/s, respectively. The waveguide, Ti(5 nm)/Au(150 nm) is also made by e-beam evaporation. Above the tri-layer samples, an 180 nm Al$_2$O$_3$ insulating layer is deposited by RF magnetron sputtering for separating the waveguide and the samples. The deposition pressure was $2\times10^{-4}$ Pa. Film crystallinity of NM layer measured by X-ray diffraction (XRD) shows in Figure S1 in supplemental information.

**Enhancement of magnetic damping constant.**

Fig. 5(a) shows rf current frequency as a function of the magnetic resonant filed. By fitting with Kittel formula, $(\omega_f/\gamma_e)^2 = \mu_0 H_{dc}(\mu_0 H_{dc} + \mu_0 M_s)$, the saturation magnetization $\mu_0 M_s$ of the Py can be derived. Fig. 5(b) shows the half width at half maximum (HWHM) as a function of rf current frequency. From the slope, we can estimate an effective magnetic damping constant $\delta_{eff}$ for Py using the following equation [30], $\Delta H = \delta_{eff}\omega_f/\gamma_e + \Delta H_0$, where $\gamma_e$ and $\Delta H_0$ are the gyromagnetic ratio of electrons and the offset of the HWHM, respectively. For Py/Cu bilayer, almost all of the injected spin current is reflected back to the Py layer without spin relaxation in Cu layer [31], because the spin diffusion length in Cu of 400 nm [30] is much larger than NM layer thickness of 20 nm. Therefore, Py/Cu bilayer sample shows the smallest slope corresponding to the smallest damping of FMR. In contrast, all of the other samples show the enhancement of damping in FMR. It implies that for Py/Ag/Bi$_2$O$_3$ and Py/Cu/Bi$_2$O$_3$, spin current is injected into the interface. On the other hand, for Py/Au/Bi$_2$O$_3$, both SOC in Au bulk and at Au/Bi$_2$O$_3$ interface contribute to the enhanced the damping of FMR. By comparison with control sample of Au/Al$_2$O$_3$, the contribution of Au/Bi$_2$O$_3$ interface for damping of FMR can be estimated as shown in Table 1.

**Estimation of spin current density.** The enhancement of the magnetic damping constant gives the spin injection efficiency known as spin mixing conductance [18],

$$g_{eff}^{\uparrow\downarrow} = \frac{4\pi M_s t_F}{g\mu_B}\left(\delta_{F/N/O} - \delta_{F/N}\right) \quad (5)$$

where $t_F$, $\delta_{F/N/O}$, and $\delta_{F/N}$ are the saturation magnetization, the thickness of Py, the damping constant for Py/NM/Bi$_2$O$_3$, and the damping constant for Py/Cu, respectively. The injected spin current density at Py/NM interface $J_s^0$ is given by [20]

$$J_s^0 = \frac{2e}{\hbar} \times \frac{\hbar g_{eff}^{\uparrow\downarrow}\gamma_e^2(\mu_0 h_{rf})^2\left[\mu_0 M_s\gamma_e + \sqrt{(\mu_0 M_s)^2\gamma_e^2 + 4\omega^2}\right]}{8\pi\delta_{F/N/O}^2[(\mu_0 M_s)^2\gamma_e^2 + 4\omega^2]} \quad (6)$$

where $h_{rf}$ and $\omega$ are the applied rf field and the angular frequency. $h_{rf}$ is determined by precession cone angle measurement developed by M. V. Costache *et al.* [33]. We measured the cone angle $\theta_c$ of the of Py(10 nm)/Al$_2$O$_3$(30 nm) bilayer sample in FMR and derived the induced $h_{rf}$ through $\theta_c = h_{rf}/2\Delta H$.

When the power of 9 GHz rf current is 20dBm, the estimated cone angle of Py/Al$_2$O$_3$ is 3.7° and the $h_{rf}$ is 9.4 Oe; the estimated spin current density $J_s^0$ of Py/Ag/Bi$_2$O$_3$, Py/Au/Al$_2$O$_3$, Py/Au/Bi$_2$O$_3$, Py/Al/Bi$_2$O$_3$, and Py/Cu/Bi$_2$O$_3$ is 13.6×10$^7$A/m$^2$, 7.7×10$^7$ A/m$^2$, 8.9×10$^7$ A/m$^2$, 9.0×10$^7$ A/m$^2$, and 11.4×10$^7$ A/m$^2$, respectively. The injected spin current $J_s^0$ at Py/NM interface propagates and exponentially decays in the NM layer. The spin current at NM/Bi$_2$O$_3$ interface is $J_{s(NM/Bi_2O_3)} = J_s^0 \times$ $\exp(-t_N/\lambda_N)$, where $t_N$ and $\lambda_N$ are the thickness and spin diffusion length of NM, respectively. For NM=Ag, Cu, Al, their $\lambda_N$ is larger than 300 nm on room temperature [32, 34, 35], which is much larger than $\lambda_N$=20nm; therefore there is almost no effect of the decay term. For NM=Au, we use $\lambda_N$=35nm from a reported value (see section 1 in supplementary information).

**First-principles calculation method.** We performed density functional calculations within the general gradient approximation [36] using OpenMX code [37], with the fully relativistic total angular momentum dependent pseudopotentials taking spin-orbit interaction (SOI) into account [38]. We adopted norm-conserving pseudopotentials with an energy cutoff of 300 Ry for charge density including the 5d, 6s and 6p-states as valence states for Bi; 2$s$ and 2$p$ for O; 3s, 3p, 3d and 4s for Cu; 4p, 4d and 5s for Ag; 5p, 5d and 6s for Au. We used 16×12×1 regular k-point mesh. The numerical pseudo atomic orbitals are used as follows: the numbers of the $s$-, $p$- and d-character orbitals are three, three and two, respectively; The cutoff radii of Bi, O, Cu, Ag and Au are 8.0, 5.0, 6.0, 7.0 and 7.0, respectively, in units of Bohr. The dipole-dipole interaction between slab models can be eliminated by the effective screening medium (ESM) method [39].

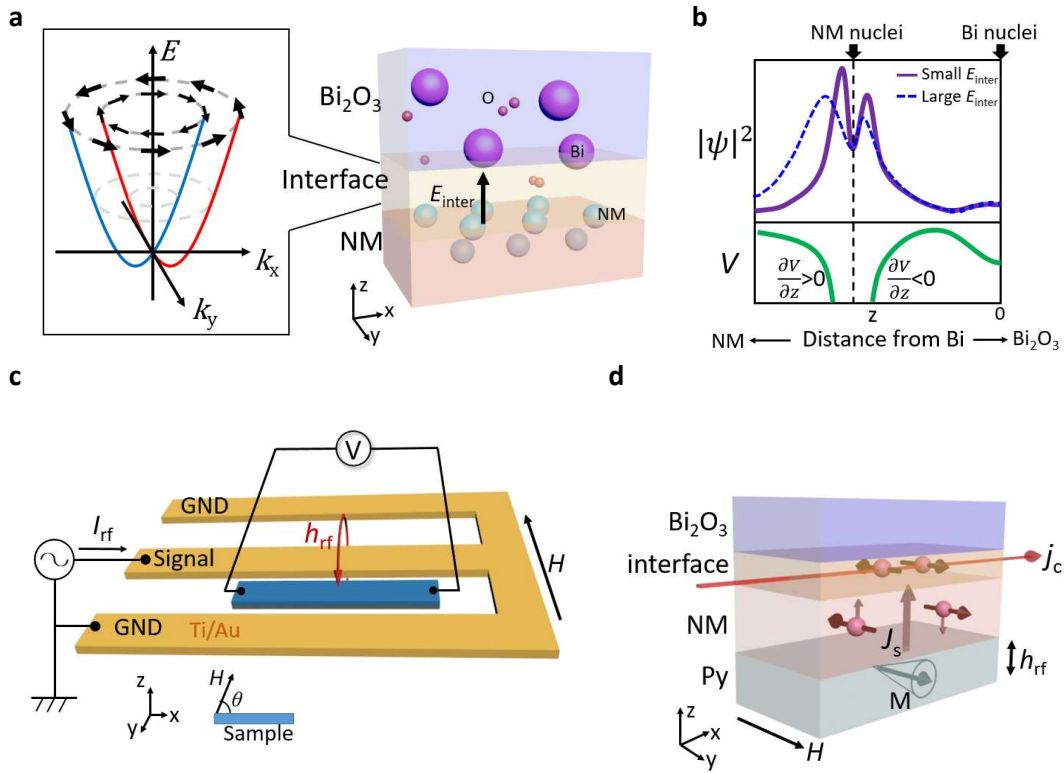

**Figure 1|Rashba spin splitting and spin-to-charge conversion in NM/Bi$_2$O$_3$ interface**
**a**, Rashba spin splitting at NM/Bi$_2$O$_3$ interface. **b**, An asymmetry distribution of $|\psi|^2$ generated by interfacial electric field $E_{\text{inter}}$. Purple line and blue line respectively show the $|\psi|^2$ under smaller and larger field $E_{\text{inter}}$. Green line show electrostatic potential $V$. **c**, Experimental setup for the spin pumping measurement. **d**, Schematic of spin-to-charge conversion at the NM/Bi$_2$O$_3$ interface. A spin current is pumped from the Py layer in resonance into the NM/Bi$_2$O$_3$ interface, and then converted to the charge current via the inverse Edelstein effect.

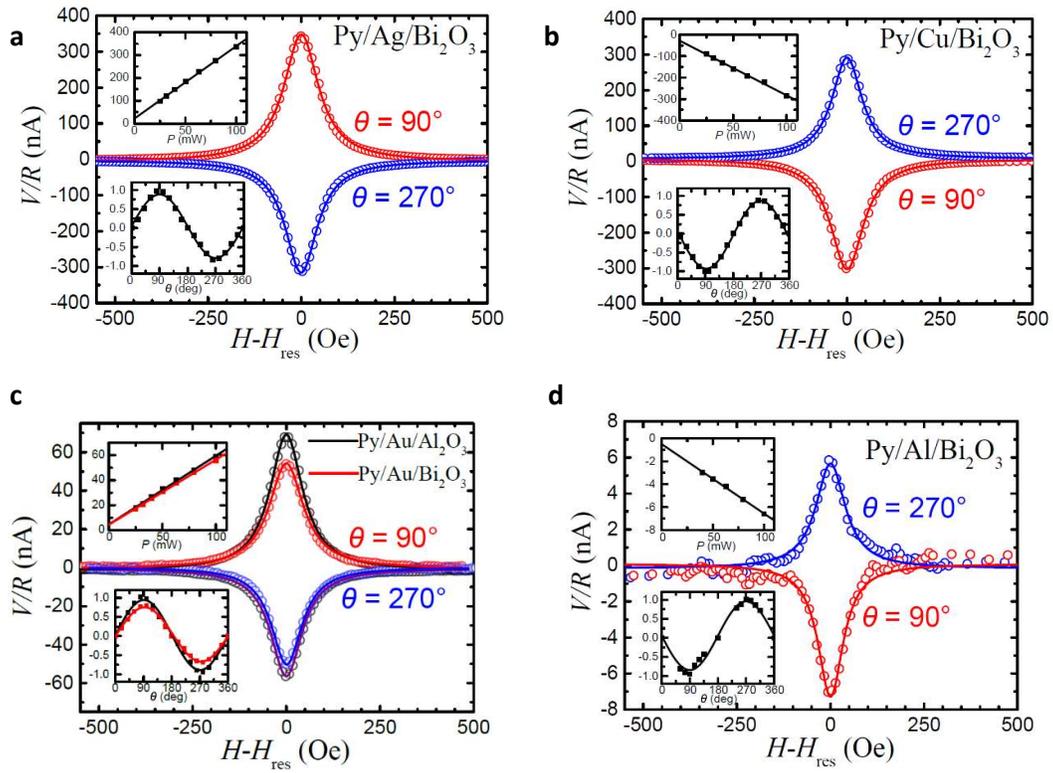

**Figure 2|Spin pumping experiment in various NM/Bi$_2$O$_3$ interface**

Detected *V/R* spectrum of **a**, Py/Ag/Bi$_2$O$_3$; **b**, Py/Cu/Bi$_2$O$_3$; **c**, Py/Au/Bi$_2$O$_3$ and Py/Au/Al$_2$O$_3$; **d**, Py/Al/Bi$_2$O$_3$. The rf power-dependence of 5 samples is shown in the upper insets, and the angle-dependence of the normalized signal *V/R* is shown in the lower insets.

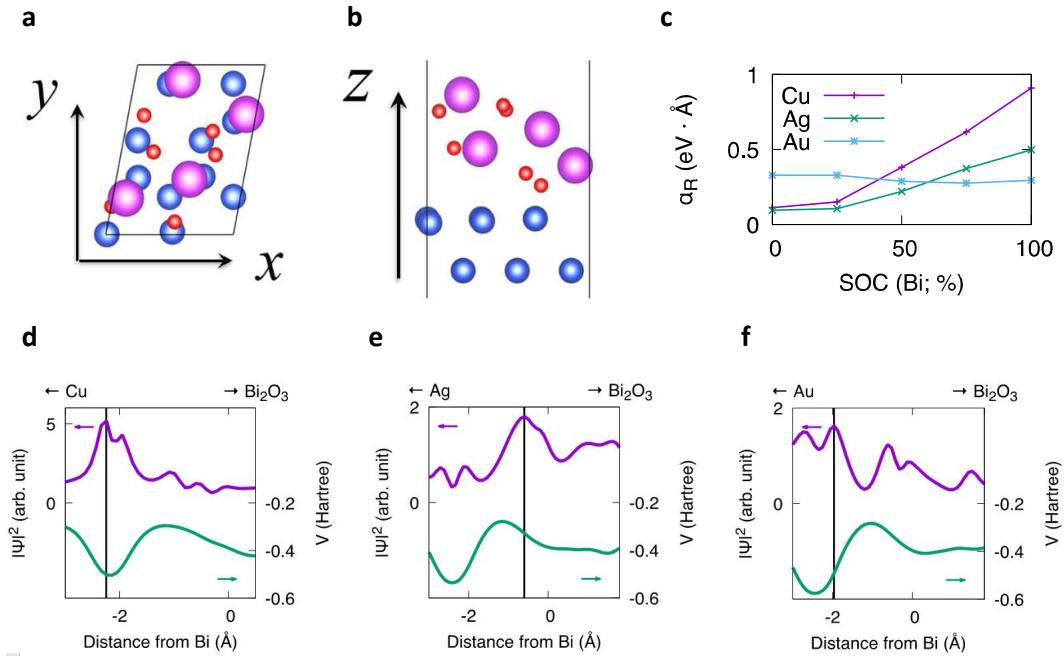

**Figure 3| Atomic structure and Charge density distribution $|\psi|^2$ of NM(111)/α-Bi$_2$O$_3$**

Atomic structure of NM(111)/α-Bi$_2$O$_3$; **a**, top view; **b**, side view. Blue, purple and red circles correspond to NM material, Bismuth and Oxygen. **c**, Strength dependence of SOC of Bi on Rashba coefficient $\alpha_R$ for NM(111)/α-Bi$_2$O$_3$. Charge density distribution $|\psi|^2$ of **d**, Cu/Bi$_2$O$_3$; **e**, Ag/Bi$_2$O$_3$; **f**, Au/Bi$_2$O$_3$. The planar averaged electrostatic potential $V$ is also shown. The origin is fixed to the position of the nearest neighbor Bi atom from top NM atom. The vertical line represents the position of the peak of $|\psi|^2$.

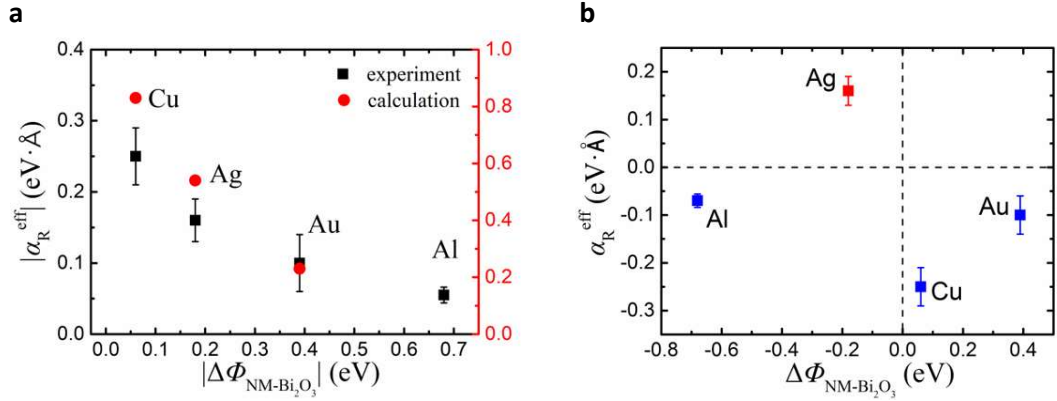

**Figure 4 | Relationship between effective Rashba parameter and work function difference**

**a**, Absolute value $|\alpha_R^{eff}|$ in various NM/Bi$_2$O$_3$ interfaces as a function of $|\Delta\Phi_{\text{NM-Bi2O3}}|$ between NM and Bi$_2$O$_3$. **b**, $\alpha_R^{eff}$ as a function of $\Delta\Phi_{\text{NM-Bi2O3}}$ between NM and Bi$_2$O$_3$.

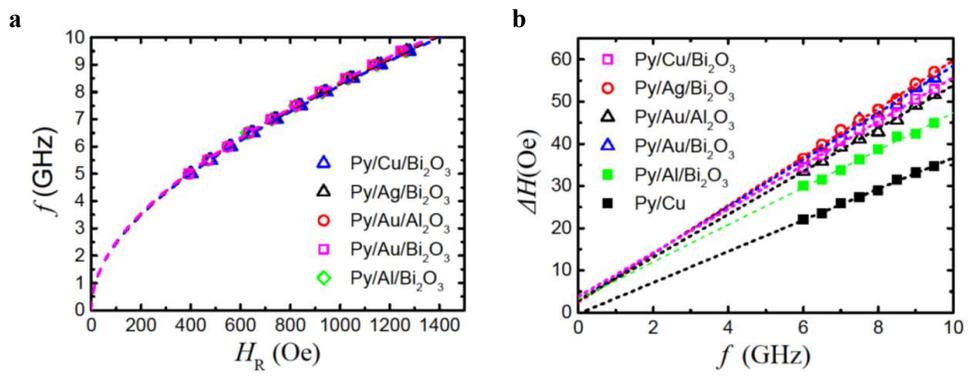

**Figure 5 | FMR measurement results in various NM/Bi$_2$O$_3$ films**

**a**, Rf current frequency as a function of the magnetic resonant filed. **b**, Half width at half maximum (HWHM) as a function of rf current frequency.

**Reference**

1. Datta, S. & Das, B. Electronic analog of the electro-optic modulator. *Appl. Phys. Lett.* **56**, 665-667 (1990).
2. Ohe, J., Yamamoto, M., Ohtsuki, T. & Nitta, J. Mesoscopic Stern–Gerlach spin filter by nonuniform spin-orbit interaction. *Phys. Rev. B* **72**, 041308 (2005).
3. Sánchez, J. *et al.* Spin-to-charge conversion using Rashba coupling at the interface between non-magnetic materials. *Nat. Commun.* **4**, 2944 (2013).
4. Zhang, W., Jungfleisch, M., Jiang, W., Pearson, J. & Hoffmann, A. Spin pumping and inverse Rashba-Edelstein effect in NiFe/Ag/Bi and NiFe/Ag/Sb. *J. Appl. Phys.* **117**, 17C727 (2015).
5. Lesne, E. *et al.* Highly efficient and tunable spin-to-charge conversion through Rashba coupling at oxide interfaces. *Nat. Mat.* **15**, 1261-1266 (2016).
6. Miron, I. *et al.* Perpendicular switching of a single ferromagnetic layer induced by in-plane current injection. *Nature* **476**, 189-193 (2011).
7. Liu, L. *et al.* Spin-Torque Switching with the Giant Spin Hall Effect of Tantalum. *Science* **336**, 555-558 (2012).
8. Karube, S., Kondou, K. & Otani, Y. Experimental observation of spin-to-charge current conversion at non-magnetic metal/$Bi_2O_3$ interfaces. *Appl. Phys. Express* **9**, 033001 (2016).
9. Kim, J. *et al.* Evaluation of bulk-interface contributions to Edelstein magnetoresistance at metal/oxide interfaces. *Phys. Rev. B* **96**, 140409(R)-1~6 (2017).
10. Puebla, J. *et al.* Direct optical observation of spin accumulation at nonmagnetic metal/oxide interface. *Appl. Phys. Lett.* **111**, 092402 (2017).
11. Petersen, L. & Hedegård, P. A simple tight-binding model of spin–orbit splitting of sp-derived surface states. *Surf. Sci.* **459**, 49-56 (2000).
12. Bentmann, H. *et al.* Spin orientation and sign of the Rashba splitting in Bi/Cu(111). *Phys. Rev. B* **84**, (2011).
13. Nagano, M., Kodama, A., Shishidou, T. & Oguchi, T. A first-principles study on the Rashba effect in surface systems. *J. Phys. Condens. Matter.* **21**, 064239 (2009).
14. Bentmann, H. & Reinert, F. Enhancing and reducing the Rashba-splitting at surfaces by adsorbates: Na and Xe on Bi/Cu(111). *New J. Phys.* **15**, 115011 (2013).
15. Niimi, Y. *et al.* Extrinsic spin Hall effects measured with lateral spin valve structures. *Phys. Rev. B* **89**, 054401 (2014).
16. Wang, H. *et al.* Scaling of Spin Hall Angle in 3d, 4d, and 5d Metals from $Y_3Fe_5O_{12}$/Metal Spin Pumping. *Phys. Rev. Lett.* **112**, 197201 (2014).
17. Mosendz, O. *et al.* Detection and quantification of inverse spin Hall effect from spin pumping in permalloy/normal metal bilayers. *Phys. Rev. B* **82**, 214403 (2010).
18. Vlaminck, V., Pearson, J., Bader, S. & Hoffmann, A. Dependence of spin-pumping spin Hall effect

14**Reference**

1. Datta, S. & Das, B. Electronic analog of the electro-optic modulator. *Appl. Phys. Lett.* **56**, 665-667 (1990).
2. Ohe, J., Yamamoto, M., Ohtsuki, T. & Nitta, J. Mesoscopic Stern–Gerlach spin filter by nonuniform spin-orbit interaction. *Phys. Rev. B* **72**, 041308 (2005).
3. Sánchez, J. *et al.* Spin-to-charge conversion using Rashba coupling at the interface between non-magnetic materials. *Nat. Commun.* **4**, 2944 (2013).
4. Zhang, W., Jungfleisch, M., Jiang, W., Pearson, J. & Hoffmann, A. Spin pumping and inverse Rashba-Edelstein effect in NiFe/Ag/Bi and NiFe/Ag/Sb. *J. Appl. Phys.* **117**, 17C727 (2015).
5. Lesne, E. *et al.* Highly efficient and tunable spin-to-charge conversion through Rashba coupling at oxide interfaces. *Nat. Mat.* **15**, 1261-1266 (2016).
6. Miron, I. *et al.* Perpendicular switching of a single ferromagnetic layer induced by in-plane current injection. *Nature* **476**, 189-193 (2011).
7. Liu, L. *et al.* Spin-Torque Switching with the Giant Spin Hall Effect of Tantalum. *Science* **336**, 555-558 (2012).
8. Karube, S., Kondou, K. & Otani, Y. Experimental observation of spin-to-charge current conversion at non-magnetic metal/$Bi_2O_3$ interfaces. *Appl. Phys. Express* **9**, 033001 (2016).
9. Kim, J. *et al.* Evaluation of bulk-interface contributions to Edelstein magnetoresistance at metal/oxide interfaces. *Phys. Rev. B* **96**, 140409(R)-1~6 (2017).
10. Puebla, J. *et al.* Direct optical observation of spin accumulation at nonmagnetic metal/oxide interface. *Appl. Phys. Lett.* **111**, 092402 (2017).
11. Petersen, L. & Hedegård, P. A simple tight-binding model of spin–orbit splitting of sp-derived surface states. *Surf. Sci.* **459**, 49-56 (2000).
12. Bentmann, H. *et al.* Spin orientation and sign of the Rashba splitting in Bi/Cu(111). *Phys. Rev. B* **84**, (2011).
13. Nagano, M., Kodama, A., Shishidou, T. & Oguchi, T. A first-principles study on the Rashba effect in surface systems. *J. Phys. Condens. Matter.* **21**, 064239 (2009).
14. Bentmann, H. & Reinert, F. Enhancing and reducing the Rashba-splitting at surfaces by adsorbates: Na and Xe on Bi/Cu(111). *New J. Phys.* **15**, 115011 (2013).
15. Niimi, Y. *et al.* Extrinsic spin Hall effects measured with lateral spin valve structures. *Phys. Rev. B* **89**, 054401 (2014).
16. Wang, H. *et al.* Scaling of Spin Hall Angle in 3d, 4d, and 5d Metals from $Y_3Fe_5O_{12}$/Metal Spin Pumping. *Phys. Rev. Lett.* **112**, 197201 (2014).
17. Mosendz, O. *et al.* Detection and quantification of inverse spin Hall effect from spin pumping in permalloy/normal metal bilayers. *Phys. Rev. B* **82**, 214403 (2010).
18. Vlaminck, V., Pearson, J., Bader, S. & Hoffmann, A. Dependence of spin-pumping spin Hall effect
14

**Acknowledgements**

This work was supported by Grant-in-Aid for Scientific Research on Innovative Area, "Nano Spin Conversion Science" (Grant No. 26103002 and No.15H01015), Grant-in-Aid for Young Scientists (B) (Grant No. JP17K14077) from MEXT. The first-principles calculation was supported in part by MEXT as a social and scientific priority issue (Creation of new functional devices and high-performance materials to support next-generation industries) to be tackled by using post-K computer (Project ID: hp160227). H.T. was supported by Japan-Taiwan Exchange Association Scholarships. S. K. was supported by Japan Society for the Promotion of Science through Program for Leading Graduate Schools (MERIT).

**Author contributions**

F. I. and Y. O. conceived the project. H. T., S. K. and K. K. designed and performed spin pumping measurement. H. T and K. K wrote the main paper. N. Y. and F. I. performed the first-principles calculation and wrote the calculation part. All authors discussed the results and commented on the manuscript.

**Competing interests**

The authors declare no competing interests.

# Supplemental information

# Clear variation of spin splitting by changing electron distribution at non-magnetic metal/Bi$_2$O$_3$ interfaces

H. Tsai[1], S. Karube[1], K. Kondou[2*], N. Yamaguchi[3], F. Ishii[4] and Y. Otani[1, 2,*)]

[1] *Institute for Solid State Physics, University of Tokyo, Kashiwa 277-8581, Japan*
[2] *Center for Emergent Matter Science, RIKEN, 2-1 Hirosawa, Wako 351-0198, Japan*
[3] *Division of Mathematical and Physical Sciences, Graduate School of Natural Science and Technology, Kanazawa University, Kanazawa 920-1192 Japan*
[4] *Faculty of Mathematics and Physics, Institute of Science and Engineering, Kanazawa University, Kanazawa 920-1192, Japan.*

*Correspondence authors: kkondou@riken.jp, yotani@issp.u-tokyo.ac.jp

Table of contents

1. **Influence of spin Hall effect in bulk**

2. **Film crystallinity**

3. **Frequency dependence of spin-to-charge conversion efficiency and Rashba parameter**

4. **First-principles calculation results and spin textures**

1. **Influence of spin Hall effect in bulk**

   When measuring the spin-to-charge (S-C) conversion at the $Bi_2O_3$ interface of Ag, Cu, and Al, the spin Hall effect of these NM materials is negligible small. However, in $Au/Bi_2O_3$ case spin Hall angle of Au is one order larger than others and makes notable contribution. For analyzing $Py/Au/Bi_2O_3$ sample, the contribution of SHE of Au and IEE in $Au/Bi_2O_3$ interface need to be separated. Firstly, we measure the spin Hall angle of Au by measuring S-C conversion in $Py/Au/Al_2O_3$ sample. By solving the spin diffusion equation with the boundary condition that spin current is zero at $Au/Al_2O_3$ interface, the spin current flowing in the Au layer is

$$J_s(y) = \frac{\sinh[(t_N - y)/\lambda_N]}{\sinh(t_N/\lambda_N)} J_s^0 \quad (S1)$$

where $t_N$, and $\lambda_N$ are the thickness of NM layer, and the spin-diffusion length of NM layer, respectively. $J_s^0$ is the spin current injected at Py/Au interface which is shown in eq. (3). Here, we use $\lambda_N = 35$ nm from a reported value [*Phys. Rev. B* **88**, 064414]. The average spin current density is $\langle J_s \rangle = \frac{1}{t_N}\int_0^{t_N} J_s(y)$ and the average charge current density in three dimension is $\langle J_c \rangle = \theta_{SH}\langle J_s \rangle$. Therefore, the spin Hall angle $\theta_{SH}$ can be calculated by

$$\langle J_c \rangle = \theta_{SH}\left(\frac{2e}{\hbar}\right)\frac{\lambda_N}{t_N}\tanh\left(\frac{t_N}{2\lambda_N}\right) J_s^0 \quad (S2)$$

As the result, $\theta_{SH}$ of Au is +0.40±0.07%, which is in a good agreement with reported value measured by spin-pumping method . The next step is considering the interface effect of $Au/Bi_2O_3$. Because some spin current is injected into the $Au/Bi_2O_3$ interface, the backflow of spin current is reduced and the injected spin current increased at Py/Au interface, i.e. $J_{s(Au/Bi_2O_3)} = J_{s(Au/Al_2O_3)} + \Delta J_s$ and $\Delta J_s > 0$. Since the backflow of spin current decays from $y = t_N$ to $y = 0$, that is $\Delta J_s(y) = \Delta J_s^0 e^{y/\lambda_N}$ and $\Delta J_s^0 = J_{s\ (Au/Bi_2O_3)}^0 - J_{s\ (Au/Al_2O_3)}^0$. The spin current in $Au/Bi_2O_3$ can be expressed as

$$J_s(y) = \frac{\sinh[(t_N - y)/\lambda_N]}{\sinh(t_N/\lambda_N)} J_{s\ (Au/Al_2O_3)}^0 + \Delta J_s^0 e^{y/\lambda_N} \quad (S3)$$

Again, the average spin current density is $\langle J_s \rangle = \frac{1}{t_N}\int_0^{t_N} J_s(y)$. We assumed that the $\theta_{SH}$ of the Au bulk in $Au/Bi_2O_3$ and $Au/Al_2O_3$ are approximately equal since the typical thickness of interface layer is only 0.4 nm. By separating the contribution of ISHE and IEE, the 3D charge current density $\langle J_c \rangle$ can be expressed as

$$\langle J_c \rangle = \langle J_s \rangle \theta_{SH} + J_{s(interface)} \times \lambda_{IEE} t_N \quad (S4)$$

and then the $\lambda_{IEE}$ of $Au/Bi_2O_3$ interface is derived.

2. **Film crystallinity**

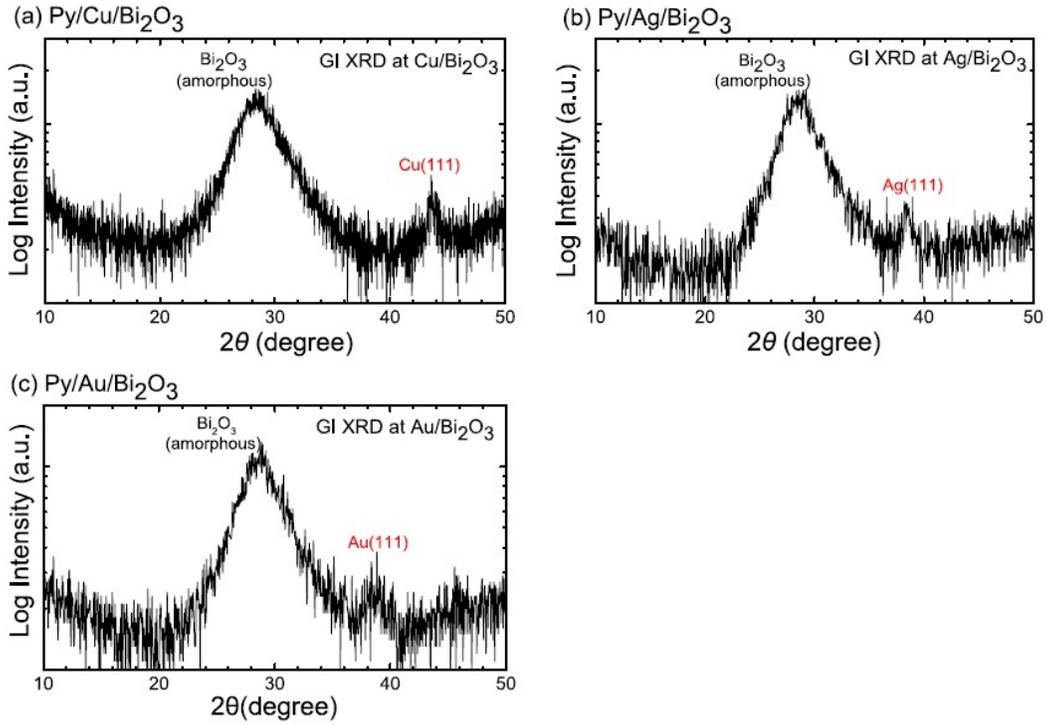

**Figure S1| X-ray diffraction results**

Grazing incident (GI) XRD spectrums at (a)Py/Cu/Bi$_2$O$_3$ (b)Py/Ag/Bi$_2$O$_3$ (c) Py/Au/Bi$_2$O$_3$ samples.

We use Grazing incident X-ray diffraction (GI XRD) to get the crystallinity information at NM/Bi$_2$O$_3$ interface of each Py/NM/Bi$_2$O$_3$ samples. For the NM layer (NM = Ag, Cu, and Au), at the interfaces Ag(111), Cu(111), and Au(111) structure are observed. These results suggest that the NM/Bi$_2$O$_3$ (NM = Ag, Cu, and Au) interfaces may have similar interface structure and therefore the strong NM dependence may not come from the crystal structure difference.

### 3. Frequency dependence of S-C conversion coefficient and effective Rashba parameter

We also investigated the frequency dependence of S-C conversion. Because the spin current generated by spin pumping in average is a dc spin current, the S-C conversion and Rashba parameter at NM/Bi$_2$O$_3$ should not depend on the frequency of rf field. As expected, by measuring the same Py/Cu/Bi$_2$O$_3$ sample at 6,7,8, and 9 GHz, $\lambda_{\text{IEE}}$ is 0.19±0.005nm and $\alpha_R^{eff}$ is 0.27±0.007 (eV·Å). The error is 2.6% which may come from the measurement and data fitting.

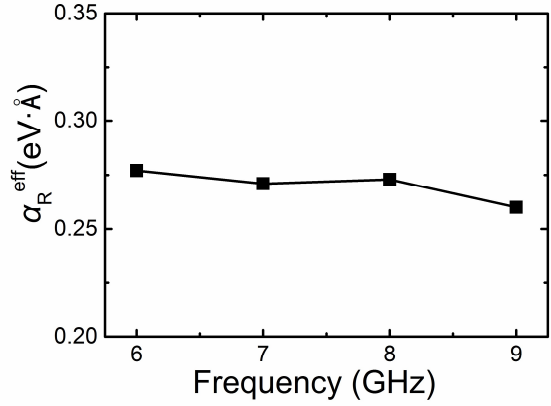

**Figure S2|Frequency dependence of Rashba parameter**

### 4. First-principles calculation results and spin textures

Fig. S3(a)-(c) shows the band structure for the NM(111)/$\alpha$-Bi$_2$O$_3$ systems, where the symmetry points ($\Gamma$, C, X) are those in the first Brillouin zone shown in Fig. S4(a). There is a free-electron-like band around C-point near the Fermi energy for each system, and its Rashba spin splitting is anisotropic. A trend in the Rashba spin splitting is corresponding to experimental one, and we obtained the Rashba coefficients $\alpha_R$ as the average of the ones along C$\Gamma$ and CX line around C-point. Our calculated $\alpha_R$ are 0.91, 0.50 and 0.29 for NM = Cu, Ag and Au, respectively, in units of eV·Å.

Fig. S4(a) shows the schematic of the first Brillouin zone of $\alpha$-Bi$_2$O$_3$. Fig. S4(b)-(d) shows the spin textures for the NM(111)/$\alpha$-Bi$_2$O$_3$ system. The anisotropic Rashba spin structures are shown for NM = Cu (Fig. S4 (b)) and for NM = Ag (Fig. S4 (c)), while the non-Rashba type spin structures are shown for NM = Au (Fig. S4 (d)). Since $\alpha$-Bi$_2$O$_3$ is monoclinic (P2$_1$/c, No. 14) and C-point is Brillouin zone-boundary, each system has no 4-fold rotational symmetry (around C-point) that makes Fermi surface and spin textures isotropic. The anisotropic Rashba spin vortices for NM = Cu and Ag are opposite to each other (e.g. The inner (outer) vortex for NM = Cu is the clockwise (anti-clockwise), while that for NM = Ag is the anti-clockwise (clockwise).), which may support our experimental result that the sign of $\alpha_R^{eff}$ in Ag/Bi$_2$O$_3$ is positive while that in Cu/Bi$_2$O$_3$ is negative. For NM = Au, there are non-Rashba type spin splitting. This may be due to strong SOC of Au. On the other hand, in the experiment, a symmetric circular spin structure was observed by angle dependence results but not an anisotropic one, because the amorphous Bi$_2$O$_3$ results in a symmetric potential in x-y plane.

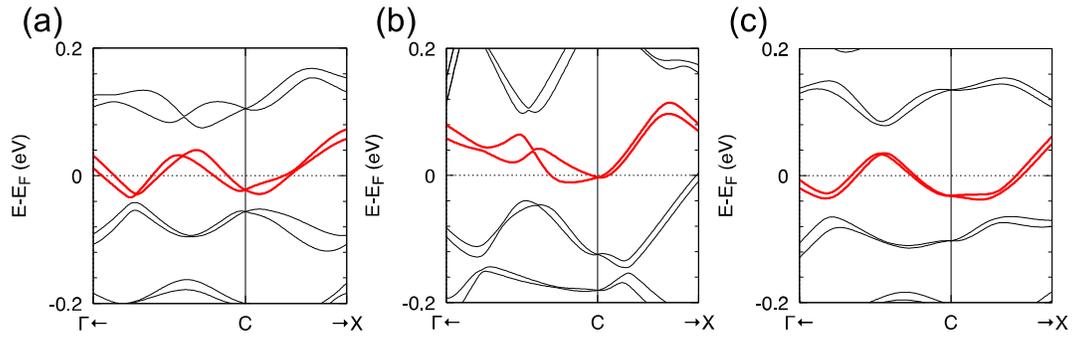

**Figure S3| Band structures for NM(111)/α-Bi$_2$O$_3$.** (a) NM = Cu; (b) NM=Ag; (c) NM=Au. The enlarged views of the band structures around C-point are shown through each path from C-point to the point dividing CΓ or CX line internally in the ratio 1:4.

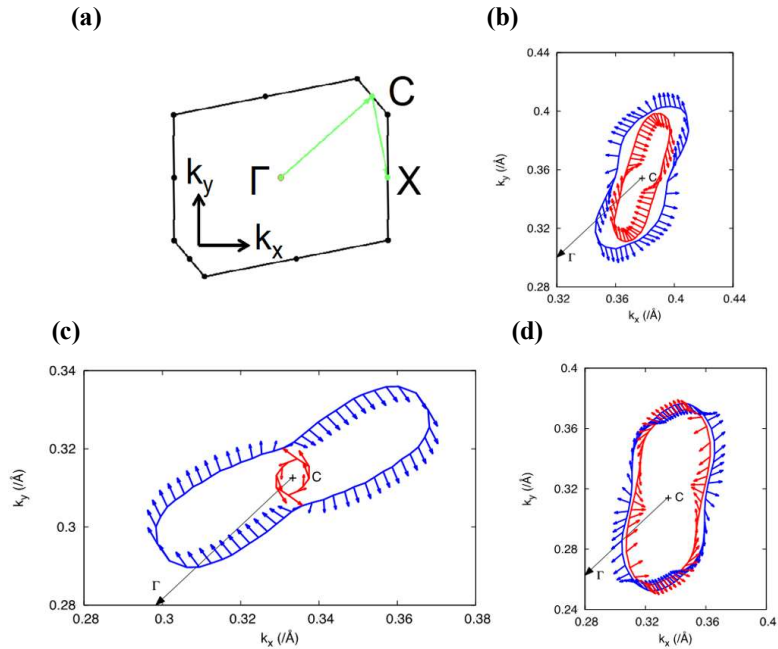

**Figure S4| Atomic structure and spin textures of NM(111)/α-Bi$_2$O$_3$**

(a) Schematic of the first Brillouin zone with high symmetry points. Spin textures of (b) Cu(111)/α-Bi$_2$O$_3$; (c) Ag(111)/α-Bi$_2$O$_3$; (d) Au(111)/α-Bi$_2$O$_3$. The black arrow shows CΓ line in the first Brillouin zone.